\documentclass{iaus}
\usepackage{graphicx}

\title[A Spizter Survey of Orion] 
{A {\it Spitzer}/IRAC Survey of the Orion Molecular Clouds}

\author[S. T. Megeath et al.] 
       {S. T. Megeath$^1$, K. M. Flaherty$^2$, J. Hora$^1$,
         L. E. Allen$^1$, G. G. Fazio$^1$, L. Hartmann$^1$,
         P. C. Myers$^1$, J. Muzerolle$^3$, J. L. Pipher$^2$, 
         N. Siegler$^3$, J.~R. Stauffer$^4$, E. Young$^3$}

\affiliation{$^1$Harvard Smithsonian Center for Astrophysics,
60 Garden St, Cambridge, MA 02138
\break email: tmegeath@cfa.harvard.edu\\[\affilskip]
$^2$Department of Physics and Astronomy, University of Rochester\\[\affilskip]
$^3$Steward Observatory, University of Arizona\\[\affilskip]
$^4${\it Spitzer} Science Center, California Institute of Technology\\[\affilskip]}

\pubyear{2005}
\volume{227}  
\pagerange{1-5}
\date{June 7, 2005}
\jname{Massive Star Birth}
\editors{R. Cesaroni, E. Churchwell, M. Felli, and C.M. Walmsley}
\begin{document}

\maketitle

\begin{abstract}

We present initial results from a survey of the Orion A and B
molecular clouds made with the InfraRed Array Camera (IRAC) onboard
the {\it Spitzer} Space Telescope.  This survey encompasses a total of
5.6 square degrees with the sensitivity to detect objects below the
hydrogen burning limit at an age of 1 Myr.  These observations cover a
number of known star forming regions, from the massive star forming
clusters in the Orion Nebula and NGC~2024, to small groups of low mass
stars in the L1641.  We combine the IRAC photometry with photometry
from the 2MASS point source catalog and use the resulting seven band
data to identify stars with infrared excesses due to dusty disks and
envelopes.  Using the presence of an infrared excess as an indicator
of youth, we show the distribution of young stars and protostars in
the two molecular clouds.  We find that roughly half of the stars are
found in dense clusters surrounding the two regions of recent massive
star formation in the Orion clouds, NGC 2024 and the Orion Nebula.


\end{abstract}

\section{Introduction}

The spatial distribution of young stars in a molecular cloud is an
important constraint on the process of star formation, providing
information on the environments in which stars and planets form and a
fossil record of the distribution of star formation sites.  Infrared
observations can readily detect young stars still embedded in their
natal gas and dust, and infrared surveys are an efficient means to
perform a census of young stars in a giant molecular cloud.  The first
large scale survey of a giant molecular cloud with an infrared array
was reported by \cite{lada1991}.  This survey showed the presence of
three large clusters and one smaller group in the Orion B cloud. Lada
et al. estimated that 58-82\% of the stars were in clusters and
groups, the major uncertainty being the number of field stars
(foreground and background) contaminating the survey.  A search for
isolated young stars outside the clusters in the Orion B cloud was
conducted by \cite{li1997} using near-IR $J$, $H$ and $K$-band
photometry.  They found that only a small percentage of the sources
showed infrared excesses in $K$-band, suggesting that the stars
outside the clusters were primarily contaminating field stars.

In contrast, the Orion A cloud contains not only the largest known
young stellar clusters within 1 kpc of the Sun (\cite[Hillenbrand \&
  Hartmann 1998]{hillen1998}; \cite[Porras et al. 2003]{porras2003};
\cite[Lada \& Lada 2003]{ladalada2003}), but also a large number of
small groups and a significant distributed population (\cite[Chen \&
  Tokunagua 1994]{chen1994}; \cite[Strom, Strom \& Merrill
  1993]{strom1993}; \cite[Allen 1996]{loriphd}).  \cite{carp2000} used
the 2MASS 2nd incremental release to survey the populations in both
the Orion A and B clouds.  By mapping the surface density of sources
and subtracting out the estimated density of contaminating field
stars, Carpenter identified nine groups and clusters in 
the clouds, and found evidence for a significant distributed
population in the Orion A cloud, but not in the Orion B cloud.  These
results suggest that most of the stars are found in the largest
clusters of the Orion A and B clouds (the Orion Nebula Cluster and the
NGC 2024 cluster, respectively).

With the goal of obtaining a census of embedded stars with disks and
envelopes in the Orion molecular clouds, we have undertaken a large
scale survey of the Orion A and B clouds with the {\it Spitzer} Space
Telescope.  We report here initial results of the survey taken with
the IRAC instrument, showing the distribution of young stars. These
observations detail the the relationship between the rich clusters,
small groups and more isolated stars in these regions.

\section{The {\it Spitzer} Orion Survey}

\begin{figure}
 \includegraphics[height=5in,width=5in]{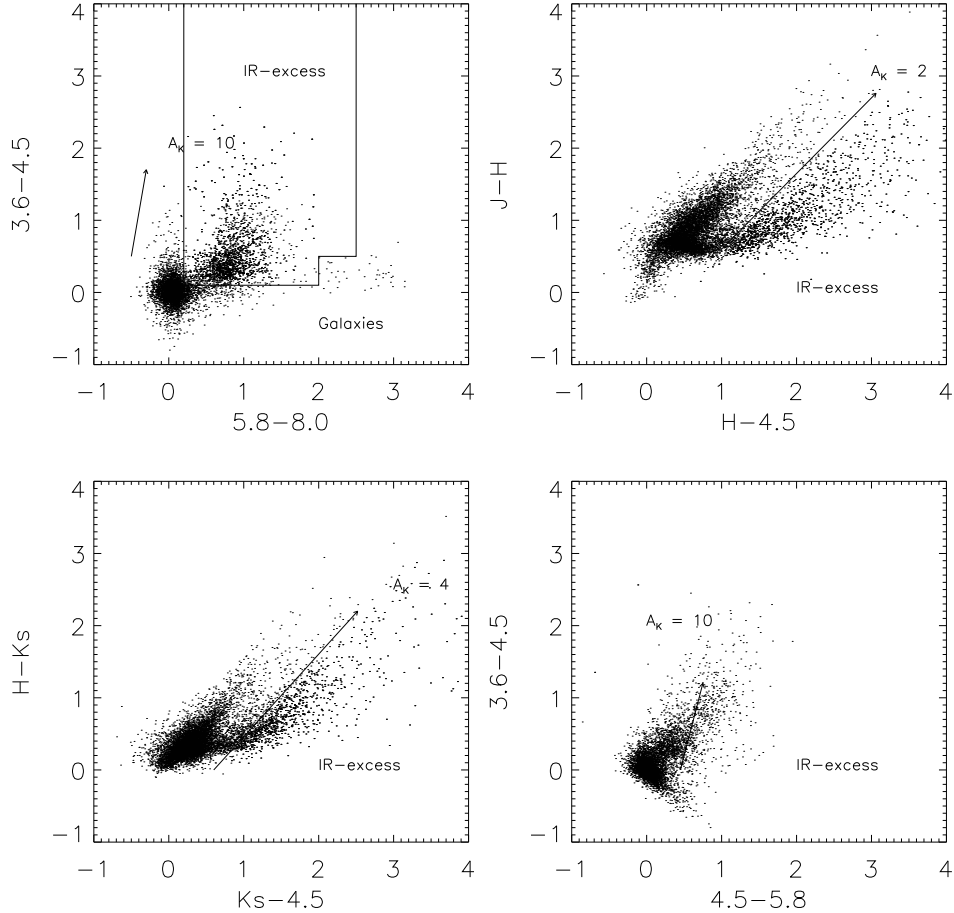}
  \caption{Color-color diagrams for the Orion A cloud, showing all sources
    with uncertainties less than 0.1~mag in the displayed wavelength bands.  
    The arrows show the direction of the extinction vector derived from
    the Orion data.  The infrared excess region of the $5.8-8.0$ vs. $3.6-4.5$
    diagram is shown. Infrared excesses were also identified from the $J-H$ vs
    $H-4.5$ and $H-K$ vs $K-4.5$ diagrams by taking points to the right of
    the reddening vectors.}
\label{fig:colorcolor}
\end{figure}

The Orion A and B molecular clouds were surveyed with the InfraRed
Array Camera (IRAC) and Multiband Imaging Photometry for Spitzer
(MIPS) as part of a collaboration between the IRAC and MIPS instrument
teams (\cite[Fazio et al. 2004]{irac}; \cite[Rieke et al. 2004]{mips}).
These data provide photometry in six wavelength bands ranging from 3
to 70~$\mu$m.  In this article, we will report solely on large scale
maps made with the IRAC instrument.  The data were taken in high
dynamic range mode, resulting in 0.6 and 12 second integration at each
position.  Four dithered integrations were obtained per position,
giving a sensitivity of 17, 16, 14.5 and 13.5~mag. in the 3.6~$\mu$m,
4.5~$\mu$m, 5.8~$\mu$m and 8.0~$\mu$m bands, respectively.  A total of
5.6 sq. degrees were covered in three fields: two fields covering the
Orion B cloud and one field coverage the Orion A cloud.  The data were
mosaicked together and a custom source finder was used to identify
sources.  Aperture photometry was performed on the identified sources
using a $2.4''$ radius aperture and a sky annulus spanning $2.4''$ to
$7.2''$.

\section{The Identification of Young Stellar Objects}

Young stars with dusty disks and infalling envelopes can be identified
by the thermal emission from the dust heated by the star.  In the
mid-infrared, the thermal emission can be detected as an excess over
the expected photospheric emission, and we refer to stars exhibiting
such an excess as infrared excess sources. We expect that more than
50\% of all young ($< 3$~Myr) stars exhibit infrared excesses
(\cite[Haisch, Lada \& Lada 2001]{haisch2001}).  Using both
theoretical models (\cite[Allen et al. 2004]{lori2004}; \cite[Whitney
  et al. 2003]{whitney2003}) and IRAC photometry of well studied young
stars and protostars in the Taurus dark clouds (\cite[Hartmann et
  al. 2005]{hartmann2005}), we have defined the IRAC colors of young
stars with disks and envelopes.  However, identification of young
stars using solely the IRAC data is limited by the lower sensitivity
of the 5.8 and 8.0~$\mu$m IRAC data to stellar photospheres.  For this
reason, we have merged the IRAC data with the 2MASS point source
catalog, and we use the $J-H$ vs. $H-[4.5]$ and $H-K$ vs. $K-[4.5]$
color-color diagrams to identify infrared excesses for young stars
which are not detected in the longer wavelength IRAC bands (Fig ~1).

The selection of sources with infrared excesses eliminates
contamination by normal field stars, but it does not eliminate
contamination due the presence of other dusty objects in the sky.
Using an IRAC survey toward a field in the constellation Bootes,
\cite{stern2005} found that AGN previously identified through optical
spectroscopy have colors similar to those of young stars with infrared
excesses.  However, the AGN are typically fainter than the young stars
in the IRAC bands. To reduce the contamination by AGN, we required
that the infrared excess sources be brighter than 14~mag. in the IRAC
3.6~$\mu$m band.  After applying this criteria, we estimate that only
42 out of the 2168 infrared excess sources in our surveyed fields are
misidentified AGNs.

\begin{figure}
{
 \includegraphics[height=3.2in, angle=-0]{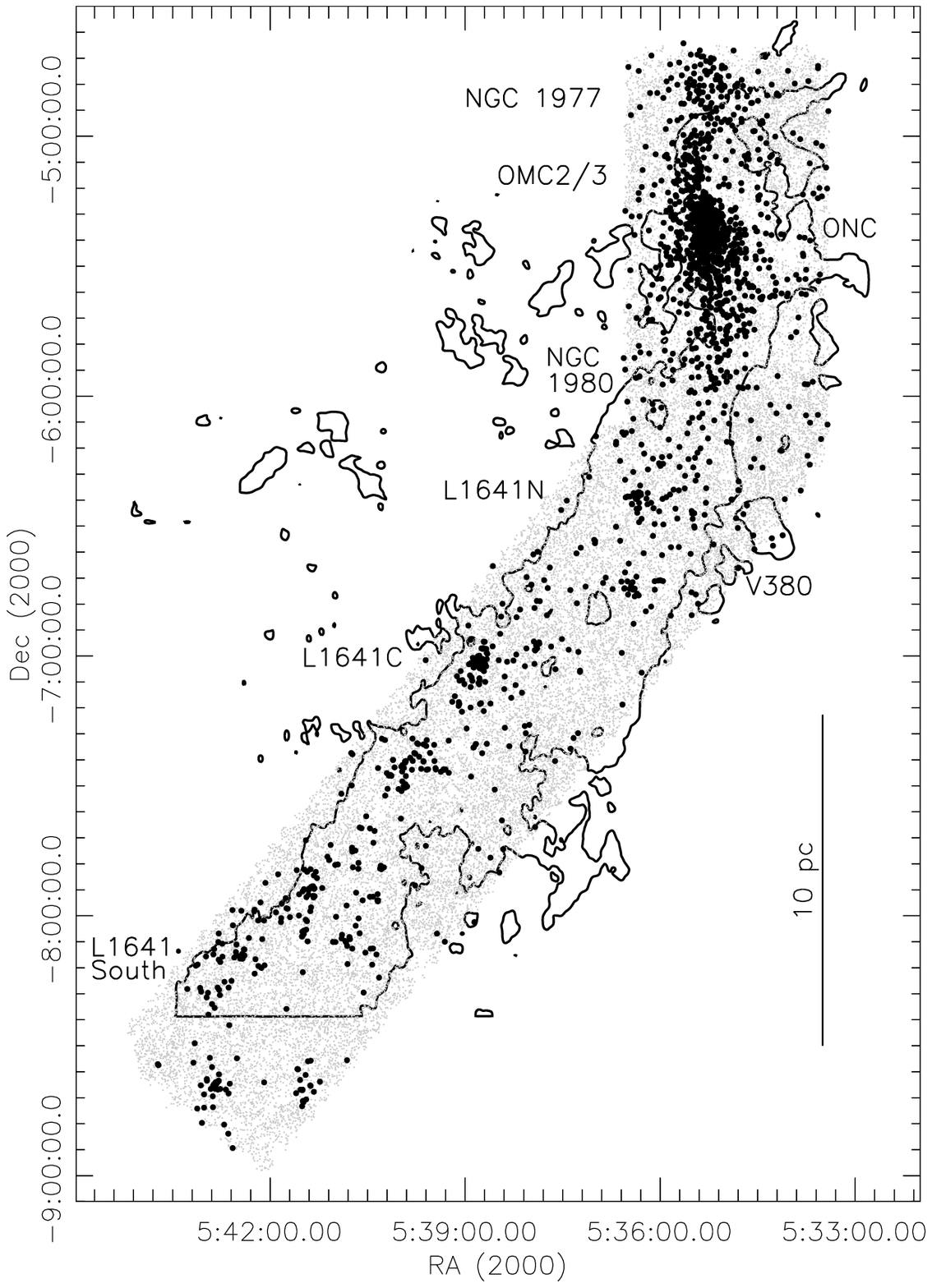}
 \includegraphics[height=3.2in, angle=-0]{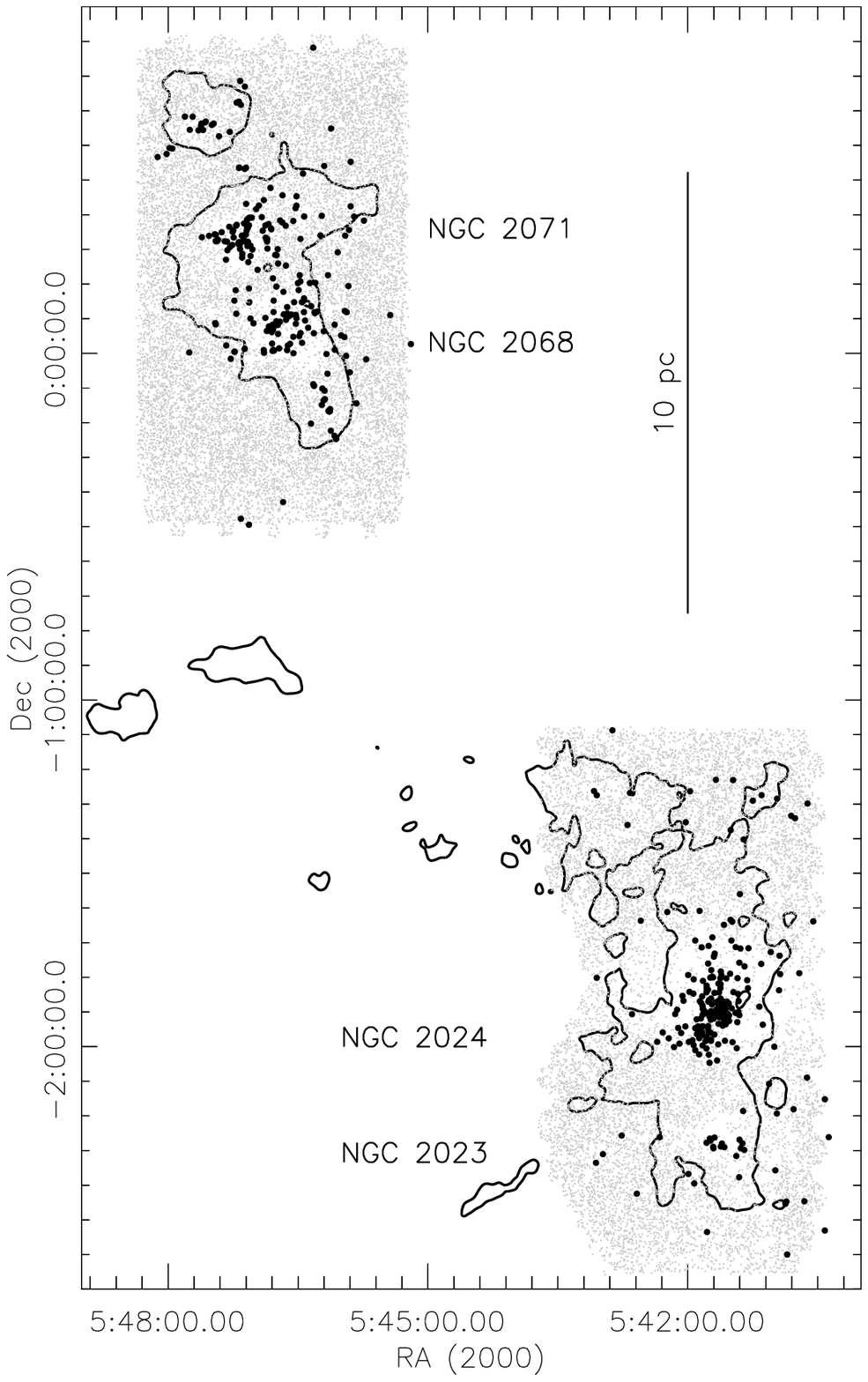}
\vfil
}
\vskip 1.0 in
  \caption{Distribution of young stars in the Orion A molecular cloud
    (left) and Orion B molecular cloud (right).  The contours show the
    velocity integrated emission in the $^{13}$CO $1 \rightarrow 0$
    transition from the Bell Labs maps (\cite[Bally et
      al. 1987]{bally1987}; \cite[Miesch \& Bally 1994]{miesch1994});
    the contour levels are 5~K~km~$s^{-1}$ and 10~K~km~$s^{-1}$ for
    the Orion A and B clouds, respectively.  The grey dots mark all
    point source detected in the 3.6~$\mu$m and 4.5~$\mu$m bands,
    while the filled circtles mark the positions of infrared excess
    sources.}
\label{fig:orionmap}
\end{figure}


\section{The Distribution of Sources in Orion}

In Fig.~2 we show the distribution of all infrared excess sources
identified in the combined IRAC and 2MASS point source catalog.  The
fields are overlaid on the velocity integrated $^{13}$CO $1
\rightarrow 0$ emission contours from the Bell Labs survey.  Two
large, centrally condensed clusters are apparent, one coincident with
the NGC 2024 region in the Orion B cloud and the other with the Orion
Nebula in the Orion A cloud.  These are the two massive star forming
regions in these clouds, and they are easily distinguished in the
figure by the density of infrared excess sources in these regions.

The Orion Nebula Cluster in Fig.~2. is part of an extended complex of
star formation extending from NGC 1977 in the north, through the Orion
Molecular Cloud (OMC)~2 and 3 regions, through M42 and M43 (the Orion
Nebula Cluster or ONC), and finally to NGC 1980 in the south.  At this
point the density of stars decreases as we transition into the L1641
dark cloud.  In L1641, we find the known stellar groups L1641 North,
V380, L1641~Center and L1641~South and an extended component which
fills in much of the space between the groups.  The stars are
concentrated toward the northern edge of the cloud where the $^{13}$CO
emission is strongest.

The two fields surveyed in the Orion B cloud show strikingly different
distributions of stars.  In the southern field containing NGC~2024 and
NGC~2023, the stars are concentrated in a single dense cluster in
NGC~2024, with a small group of stars associated with NGC~2023 and a
small number of isolated stars.  In the northern field containing
NGC~2068 and NGC~2071, we find that the two clusters are part of a
more extended distribution of stars spanning from NGC~2071 through
NGC~2068 and to the chain of protostars found in the LBS 23 core -
which includes HH24 and McNeil's nebula.  In addition, a small group
is seen in the northernmost part of the Orion B cloud, this group is
outside the region surveyed by Lada et al. 1992.

\begin{figure}
 \includegraphics[height=4.in,width=5in]{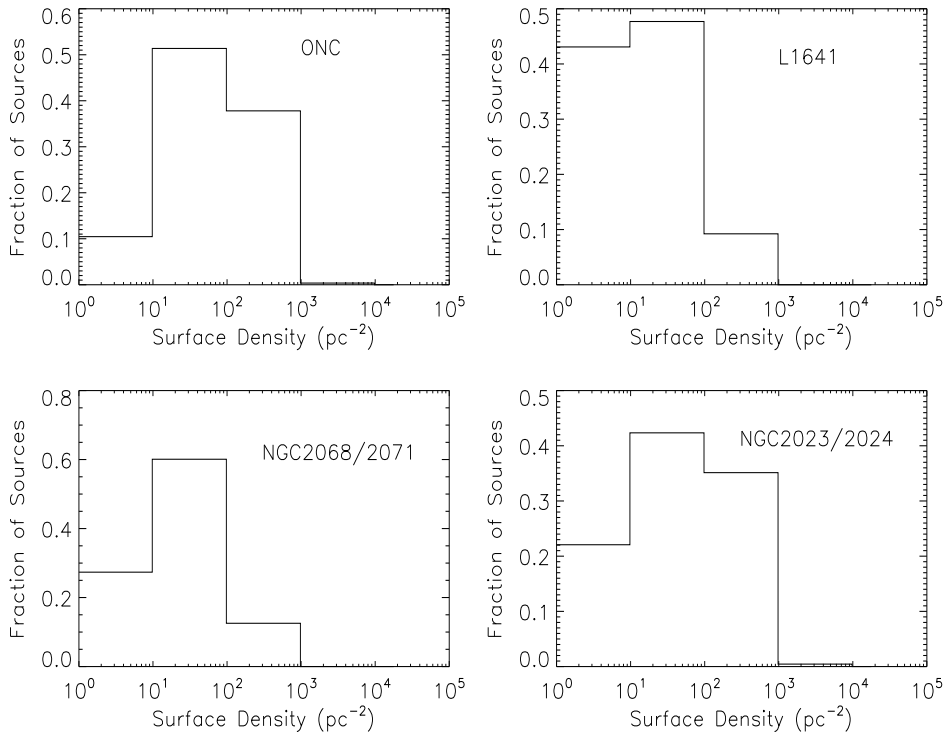}
  \caption{Histograms of the local density of infrared excess
    sources.  For each infrared excess source, the local stellar
    density was determined using the nearest neighbour method.  The
    resulting histograms of the densities differ between the four
    sub-regions of the Orion A and B clouds.  The L1641 and ONC
    regions are the regions of the Orion A cloud south and north of
    $Dec = -6^o$, respectively.}
\label{fig:nearest}
\end{figure}

\section{Clustered vs. Distributed Star Formation}

The different spatial distributions of young stars observed in the
clouds is apparent in histograms of the local surface density of
infrared excess sources.  We calculate the local surface density for
each infrared excess source by finding the distance to the 4th nearest
infrared excess source, $r$, and then equating the local surface
density to $5/\pi r^2$.  Fig~4 shows the distributions of nearest
neighbour densities for the two maps in the Orion B cloud and for two
sub-regions of the Orion A cloud map.  In NGC~2068/2071 and L1641, $<
10\%$ of the stars are in regions with stellar densities $>
100$~stars~pc$^{-2}$, while in NGC~2023/2024 and the Orion Nebula
Cluster (ONC), more than $> 30\%$ of the stars have local densities
$>100$~stars~pc$^{-2}$.  In the ONC, this fraction is probably much
larger since the data is incomplete in the center of the Orion Nebula
where the stellar densities are the highest.  NGC~2068/2071 shows a
strong peak between 10 and 100~stars~pc$^{-2}$, suggesting a more
uniform distribution of stars compared with the other regions.
Interestingly, the regions showing dense, centrally condensed clusters
are the two regions of the embedded molecular cloud population closest
to the older Orion 1C association (see article by Bally et al. in this
volume), suggesting that the rate and/or density of star formation may be
enhanced by the impact of OB stars on the molecular clouds.

One of the goals of this survey is to determine the relative importance of
clustered and isolated star formation in the Orion molecular clouds;
however, distinguishing the clustered and distributed stars can be
problematic.  Only in the NGC~2023/2024 cloud are the clustered and
distributed populations distinct, while in the other regions the
clusters and groups appear to be density peaks in more extended
populations.  As a first step, we calculate the number of stars in the
two large, dense clusters in the Orion and NGC~2024 regions relative
to that in the remaining, lower density regions.  We note that in the
Orion~B cloud, there are roughly the same number of infrared excess
sources in the predominately clustered population of the NGC~2023/2024
sub-cloud as in the lower stellar density, more distributed population
in the NGC~2068/2071 sub-cloud (239 vs. 229, respectively).  In the
Orion~A cloud, we find 607 infrared excess sources in L1641, while
north of $Dec = -6^o$, we find 1093 sources, 487 of which appear to be
part of the ONC, and the remaining 606 stars are associated with
OMC2/3, NGC 1977, NGC 1980, and a more distributed component seen
around the cluster.  We estimate that we may be missing as many as 700
stars in the inner $5' \times 5'$ center of the Orion Nebula
(\cite[Muench et al. 1998]{gus1998}).  In total, we estimate that
there may be as many as 1200 stars with infrared excesses in the ONC
compared with 1213 in the regions surrounding the ONC and in L1641.
These preliminary results suggest that roughly half of the stars are
in the two dense, rich, centrally concentrated clusters, while the
other half are found in lower density environments.

\begin{acknowledgments}
This work is based in part on observations made with the Spitzer Space
Telescope, which is operated by the Jet Propulsion Laboratory,
California Institute of Technology, under NASA contract 1407. Support
for this work was provided by NASA through contract 1256790 issued by
JPL/Caltech.
\end{acknowledgments}

\end{document}